\documentclass[aps,prl,twocolumn,groupedaddress]{revtex4}
\usepackage{amsmath}
\usepackage{amssymb}
\usepackage{graphicx}

\begin{document}

\title{Spin-flip scattering at quantum Hall transition}
\author{V. Kagalovsky$^{1,3}$ and A. L. Chudnovskiy$^{2,3}$} 
\affiliation{$^1$ Sami Shamoon College of Engineering, 
Bialik/Basel St., Beer-Sheva 84100, Israel \\
$^2$I. Institut f\"ur Theoretische Physik, Universit\"at Hamburg,
Jungiusstr 9, D-20355 Hamburg, Germany \\ 
$^3$Max-Planck-Institut f\"ur Physik komplexer Systeme
N\"othnitzer Str. 38,  01187 Dresden, Germany}

\date{\today}

\begin{abstract} 
We formulate a generalized Chalker--Coddington network model that describes the effect of nuclear spins on the two-dimensional electron gas in the quantum Hall regime. We find exact analytical expression for the transmission coefficient of a charged particle through a saddle point potential in presence of perpendicular magnetic field that takes into account spin-flip processes.   
Spin-flip scattering creates a metallic state in a finite range around the critical energy of quantum Hall transition. As a result we find that the usual insulating phases with Hall conductance $\sigma_{xy}=0, 1, 2$ are separated by novel metallic phases. 
\end{abstract}


\maketitle

Transition between the quantized values of Hall conductance in integer quantum Hall effect represents a profound example of electronic delocalization. Properties of this transition remain a subject of intensive theoretical and experimental investigations since the discovery of the integer quantum Hall effect (QHE) 
\cite{vonKlitzing}.   Theory of the delocalization transition in the QHE  predicts the existence of a single delocalized critical electronic state at the center of  Landau level (LL)   \cite{Khmelnitskii}. Other states of a given LL are localized by disorder due to the Anderson localization phenomenon. 
Effective theoretical treatment of QHE delocalization transition is provided in frame of the Chalker--Coddington (CC) network model \cite{CC}. In this model electrons
move along unidirectional links that form closed loops in analogy with semiclassical motion on contours of constant potential. Scattering between links is allowed at nodes, in analogy with tunneling through saddle-point potentials in the
semiclassical model.  
In Ref. \cite{Shapiro} B. Shapiro  pointed out that a convenient modeling of disordered systems can be given by networks of unitary random scattering matrices that correspond to  the symmetries of the Hamiltonian. In this way, 
subsequent generalizations of CC model allowed to include spin degree of freedom 
\cite{CL} and to describe systems belonging to various symmetry classes \cite{KagPRL99,KagPRB01}.  

 In this paper we study the effect of spin-flip scattering by magnetic nuclei on the QHE transition. The spin-flip scattering by nuclei mixes two components of electron spin, and results in the mutual influence of electron and nuclear spins. The interplay of electron and nuclear spin degrees of freedom has been investigated earlier, focusing on measurement of nuclear-spin-lattice relaxation mediated by hyperfine interaction with two-dimensional electron system in QH regime \cite{vonKlitz,Vagner}. In contradistinction, we concentrate on the effect of nuclear spins on the electron motion and resulting changes in the QH phase diagram. 
 
\begin{figure}
\includegraphics[width=8cm,height=7cm,angle=0]{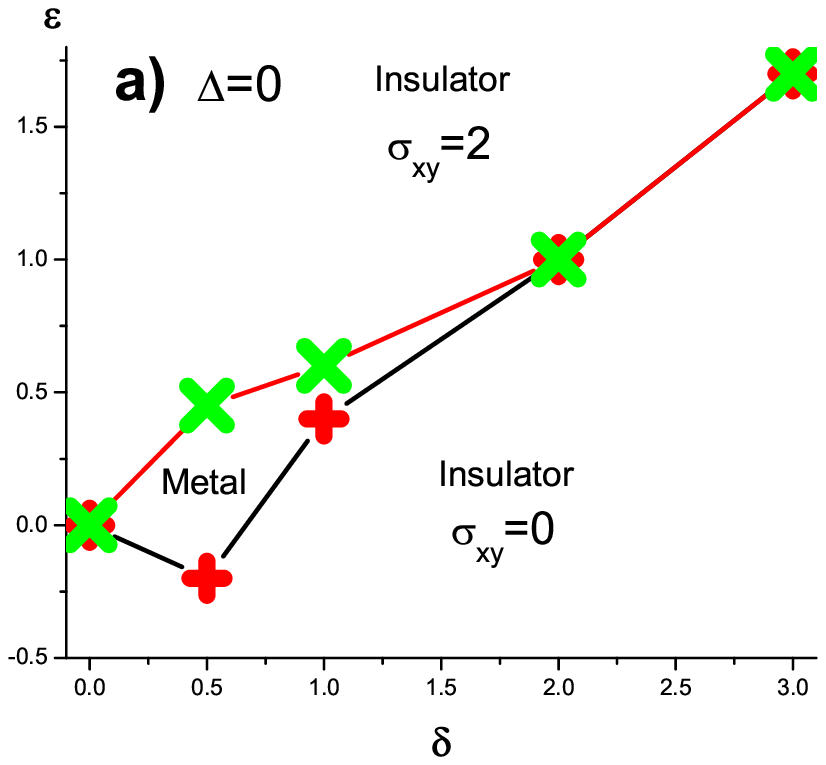}
\vskip -0.5cm
\includegraphics[width=8cm,height=7cm,angle=0]{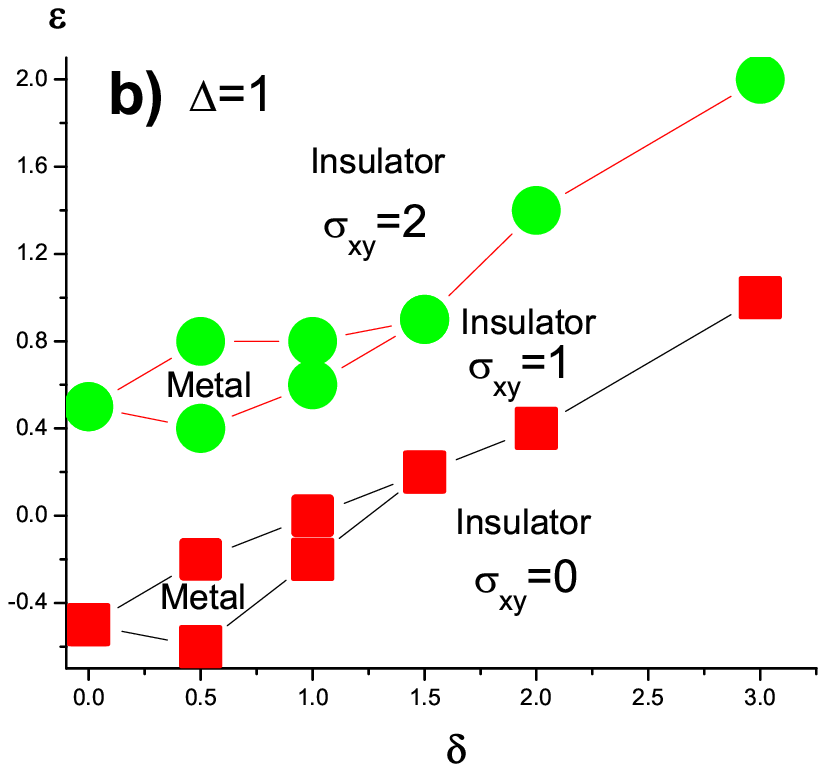}%
\vskip -0.5cm
\caption{(Color online) a) Phase diagram at zero Zeeman splitting.
b) Phase diagram with Zeeman splitting.  
\label{fig-Phases}}
\end{figure}

 We adopt the model of point-like exchange interaction between nuclear and electron spins $H_{\mathrm{int}}= J {\bf I}\cdot{\bf s}$, where ${\bf I}$ and ${\bf s}$ denote the spins of the nucleus and of the electron respectively. Throughout the paper we assume spin-1/2 nuclei.  In the absence of spin-flip scattering there are two Zeeman-split critical energies for each Landau level, where the QH delocalization transition takes place.  We find that the spin-flip scattering results in the appearance of a finite region of delocalized states around the critical QHE states. The energetical width of the delocalized region depends on the  strength of the exchange interaction relative to the Zeeman splitting in a nonmonotonous way.  It increases with the exchange strength for weak exchange, reaches its maximum, and then shrinks back to two critical states at very large exchange interaction. Our results are summarized in phase diagram shown in Fig. \ref{fig-Phases}. To obtain this phase diagram, we constructed an effective network model based on the famous CC model for the  QH transition. 
 
\begin{figure}
\includegraphics[width=8cm,height=11cm,angle=0]{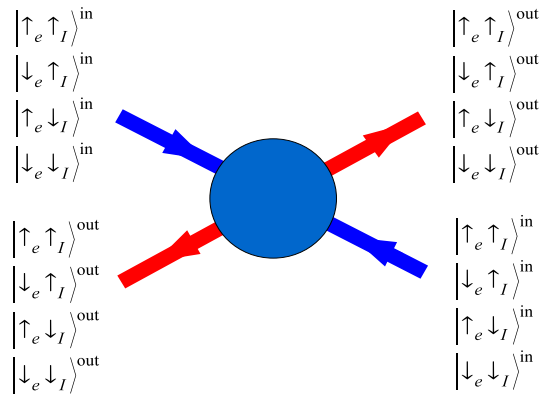}%
\vskip -3.cm
\caption{(Color online) Incoming and outgoing states at a single node.   
\label{fig-Node}}
\end{figure}

 Exact description of quantum Hall transition that takes into account scattering of electrons by impurity spins requires the knowledge of a many particle wave function 
$\Psi(t, {\bf r}, {\bf R}_1, ... {\bf R}_N )$, where ${\bf r}$ denotes the coordinate of the mobile electron and ${\bf R}_i, i=1,..., N$ are the coordinates of nuclear   spins. It follows that a formal CC model has to use  scattering matrices for the many particle wave function introduced above. However, the short range of interaction between the electron and a nuclear spin 
allows for a series of simplifications  that lead to an effective description in frame of CC model formulated for the reduced two-particle wave function. First, let us notice that far from the saddle points of disorder potential, the energies of spin-up and spin-down electrons are different due to Zeeman splitting, and the elastic spin-flip scattering is suppressed. The energy conservation can however be respected near the saddle-points \cite{KagVag07}. Therefore, in terms of CC model, the spin-flip scattering is effective only at the nodes of the network.  
Second, the act of scattering in each node involves only the coordinates of the electron and a single nucleus that is located in the node. Considered alone, this scattering event can be described in terms of the scattering matrix for the two particle wave function $\Psi_{i}(t, {\bf r}, {\bf R}_i)$, where $\bf r$ is the coordinate of the electron and ${\bf R}_i$ is the coordinate of the nucleus in the node $i$. 
\begin{equation}
\Psi^{\mathrm{out}}_{i}(t, {\bf r}, {\bf R}_i)=\hat{S} \Psi^{\mathrm{in}}_{i}(t, {\bf r}, {\bf R}_i)
\label{S2}
\end{equation}
The explicit expression for the scattering matrix elements is obtained along the lines   of Ref. \cite{FH} generalized on the problem with spin-flip scattering. 
 We assume, that the scattering at saddle points of the potential (that correspond to the nodes in CC model) is accompanied by the interaction with nuclear spin, and hence allows for spin flips. 
Thereby, the states $\vert\uparrow_e\uparrow_I\rangle$ and $\vert\downarrow_e\downarrow_I\rangle$ (where indices $e, I$ correspond to the spin state of electron and nucleus respectively) cannot flip the spin because of conservation of the total angular momentum by scattering. The scattering matrix for these two states can be obtained directly from the expression of Ref. \cite{FH} by shifting the hight of the saddle point by Zeeman energy of the electron. Here we neglect Zeeman energy of  a nucleus, which does not change the results of this paper qualitatively.  We note in passing, that the model that takes into account the Zeeman energy of the localized spin scatterers can also be applied to the scattering by localized magnetic impurities as it takes place, for example, in semimagnetic semiconductors 
\cite{semicond}. 
The states $\vert \uparrow_e\downarrow_I\rangle$ and $\vert \downarrow_e\uparrow_I\rangle$ can be mixed in the spin-flip process. To obtain the transmission coefficients for those states, we find the eigenstates and eigenenergies of the  Hamiltonian $H=H_0+H_{\mathrm{int}}+H_{\mathrm{Z}}$, where $H_0$ is the Hamiltonian describing the motion of the electron in the scalar saddle point potential \cite{FH}, and $H_{\mathrm{Z}}$ describes the Zeeman energy. 
We choose the basis of singlet and triplet states that are the eigenstates of  the interaction part of the Hamiltonian. The two-dimensional Hilbert space for the spin-flip scattering problem is then formed by the states  $\Psi_{0,0}=\frac{1}{\sqrt{2}}\left(\vert\uparrow_e, \downarrow_N\rangle 
+\vert\downarrow_e, \uparrow_N\rangle\right)$ and  $\Psi_{10}=\frac{1}{\sqrt{2}}\left(\vert\uparrow_e, \downarrow_N\rangle 
-\vert\downarrow_e, \uparrow_N\rangle\right)$. Similarly to Ref. \cite{FH} we seek the solution of the Schr\"odinger equation in the form  $\Phi_n(X,s)=\psi_n(s)\Phi_n(X)$, where  $\psi_n(s)$ and $\Phi_n(X)$ describe the cyclotron motion and the motion of the guiding center respectively, and $n$ denotes the number of Landau level(LL).   Explicit expression for $\psi_n(s)$ is provided in Ref. \cite{FH}. Without the loss of generality we restrict the further consideration to the lowest Landau level and suppress the Landau level index.  The equation for the guiding center wave function in the basis $\Phi=\left(\Psi_{0,0}, \Psi_{1,0}\right)^T$ is given by 
\begin{eqnarray}
\nonumber &&
\left\{\hat{H}_0 {\bf 1}_2 + J\left(
\begin{array}{cc}
-3/4 & 0 \\
0 & 1/4
\end{array}
\right) 
-\frac{B_0}{2}\left(
\begin{array}{cc}
0 & 1 \\
1 & 0
\end{array}
\right)
\right\} \Phi_n(X) \\
&&
=\mathcal{E} \Phi_n(X).     
\label{Hamiltonian}
\end{eqnarray}
We seek for the solution of Eq. (\ref{Hamiltonian}) in the form 
\begin{equation}
\Phi(X)=\phi(X)\left(
\begin{array}{c}
\zeta_1 \\
\zeta_2
\end{array}
\right),
\label{Ansatz} 
\end{equation}
where $\phi(X)$ describes the spatial dependence of the wave function, and $(\zeta_1, \zeta_2)^T$ is its spinor part. 
Let $\phi(X)$ be the solution of the Schr\"odinger equation  
\begin{equation}
\hat{H}_0\phi(X)=\tilde{E}\phi(X). 
\label{tildeE}
\end{equation} 
Diagonalization of the spinor part of Eq. (\ref{Hamiltonian}) for a fixed $\tilde{E}$ leads to the two eigenvectors and corresponding eigenergies
\begin{equation}
\mathcal{E}_{1,2}=\tilde{E}-\frac{1}{4}J\pm \frac{1}{2}\sqrt{J^2+B^2}. 
\label{E1,E2}
\end{equation}
The energies $\mathcal{E}_i$ correspond to the total energies of the two-particle states. By the energy conservation, both energies are to be equal to the energy of the incoming electron, $\mathcal{E}_1=\mathcal{E}_2=E$. This leads to the {\em two different values} for the energy $\tilde{E}$ that determines the effective height of the potential barrier  
\begin{equation}
\tilde{E}_{1,2}=E+\frac{1}{4}J\mp\frac{1}{2}\sqrt{J^2+B_0^2}.
\label{tildeE_12}
\end{equation}
Following \cite{FH}, we introduce the dimensionless measure of energy $\epsilon=(E+\frac{1}{4}J)/E_1$, where $E_1$ is the energetic parameter characterizing the form of the saddle point potential. Furthermore, we denote the dimensionless  strength of the Zeeman coupling as $\Delta=B/E_1$, and the relative exchange strength as $\delta=J/E_1$. 
Therefore, there are two solutions of Eq. (\ref{Hamiltonian}) corresponding to the energy $E$ of the incoming electron, $\left\vert \Phi_{1,2}(X)\right\rangle=\phi(\epsilon_{1,2},X)\left\vert \mathrm{spin}_{1,2}\right\rangle$, where 
$\epsilon_{1,2}=\epsilon\mp(1/2)\sqrt{\Delta^2+\delta^2}$,  and $\phi(\epsilon_{1,2},X)$ is the solution of 
Eq. (\ref{tildeE}) for the energy $\tilde{E}_{1,2}$ respectively. The explicit form of 
$\phi_n(\epsilon_{1,2},X)$ is given in Ref. \cite{FH}.   
The spin parts of the wave functions are given by 
\begin{equation}
\vert \mathrm{spin}_{1,2} \rangle= \frac{\left(\mathcal{D}+\delta-
\Delta\right)\vert\uparrow_e\downarrow_N\rangle 
\pm\left(\Delta+\delta+\mathcal{D}\right)\vert\downarrow_e\uparrow_N\rangle}{2\left[\mathcal{D}^2 + \mathcal{D}\delta\right]^{1/2}},   
\label{spin_i}
\end{equation} 
where $\mathcal{D}\equiv \sqrt{\Delta^2+\delta^2}$.
The energies $\epsilon_{1,2}$ together with the energies 
$\epsilon_{\uparrow,\downarrow}=\epsilon\pm \Delta/2-\delta/2$ of the states 
$\left\vert\uparrow_e\uparrow_I\right\rangle$ and $\left\vert \downarrow_e\downarrow_I\right\rangle$  determine the transmission coefficients $t(\epsilon_i)$ defined below. From Eq. (\ref{spin_i}), we obtain the transmission coefficients for the scattering with and without spin flip: 
\begin{eqnarray}
\nonumber 
^{\mathrm{out}}\langle\downarrow_e\uparrow_N\vert \uparrow_e\downarrow_N\rangle^{\mathrm{in}}=^{\mathrm{out}}\langle\uparrow_e\downarrow_N\vert \downarrow_e\uparrow_N\rangle^{\mathrm{in}}\\
=\frac{\delta}{2\mathcal{D}} \left[t(\epsilon_1)-t(\epsilon_2)\right], 
\label{flip}\\
^{\mathrm{out}}\langle\downarrow_e\uparrow_N\vert \downarrow_e\uparrow_N\rangle^{\mathrm{in}}=\frac{\mathcal{D}+\Delta}{2\mathcal{D}}t(\epsilon_1)
+\frac{\mathcal{D}-\Delta}{2\mathcal{D}}t(\epsilon_2), 
\label{no-flip1}\\
^{\mathrm{out}}\langle\uparrow_e\downarrow_N\vert \uparrow_e\downarrow_N\rangle^{\mathrm{in}}=\frac{\mathcal{D}-\Delta}{2\mathcal{D}}t(\epsilon_1)
+\frac{\mathcal{D}+\Delta}{2\mathcal{D}}t(\epsilon_2), 
\label{no-flip2} \\
^{\mathrm{out}}\langle\uparrow_e\uparrow_N 
\vert \uparrow_e\uparrow_N\rangle^{\mathrm{in}}=t(\epsilon_{\uparrow}), 
\label{upup}\\
^{\mathrm{out}}\langle\downarrow_e\downarrow_N 
\vert \downarrow_e\downarrow_N\rangle^{\mathrm{in}}=t(\epsilon_{\downarrow}),
\label{downdown}  
\end{eqnarray}
where from Ref. \cite{FH} 
\begin{equation}
t(\epsilon_i)=\frac{1}{\sqrt{1+\exp(-\pi\epsilon_i)}}. 
\label{t-FH}
\end{equation}
The elements for reflection are given by the exchange $t(\epsilon_i)\rightarrow r(\epsilon_i)=\sqrt{1-t^2(\epsilon_i)}$ everywhere.  
Using the expressions for transmission and reflection coefficients, we construct the scattering matrix in the node relating the incoming and outgoing waves (see Fig. \ref{fig-Node}). From that we derive the transfer matrix relating the waves on the left and on the right sides of the node. The system is, on average, invariant under $90^{\circ}$ rotation if at the next neighbor node the transmission and reflection (of each channel) are interchanged, i.e. $\epsilon_i \rightarrow -\epsilon_i$ \cite{CC}. 

In the standard formulation of the CC network model, the outgoing wave after the scattering at the node $i$ and acquiring a random phase on the link plays the role of the incoming wave for the scattering by the next node  (let's number it with $i+1$). The latter allows to get the information about the propagation through the network by dividing the network into slices and  multiplying the transfer matrices slice by slice. Such a direct approach is inapplicable if one replaces the exact many-particle scattering matrix by the two-particle scattering matrix though. Namely, the outgoing two-particle wave function from the node $i$ contains no information about the phase and the spin direction of the nucleus in the node $i+1$. This is in contradiction with the direct multiplication of the transfer matrices that would automatically make the outgoing state of the nucleus $i$ equal to the incoming state of the nucleus $i+1$. 
 Therefore, in order to use the two-particle scattering matrix, one has to mimic the loss of the information about the state of the nucleus $i$ before the scattering by the nucleus $i+1$. For example, the outgoing two-particle amplitude for the spin-up electron contains two contributions, $\vert \uparrow_e \uparrow_I\rangle$ and $\vert \uparrow_e \downarrow_I\rangle$.  In order to erase the memory about the state of the previous nuclear spin, we allow those two components to be exchanged randomly before being scattered by the next node. Similarly we exchange the components  $\vert \downarrow_e \uparrow_I\rangle$ and $\vert \downarrow_e \downarrow_I\rangle$. Effectively, we introduce  $\sigma^x$ Pauli matrix acting randomly with probability 1/2 on each link on the state of the nucleus.  This mechanism can also be  interpreted as a spin-orbit scattering acting on nuclear spins only. Therefore, we arrive at the version of CC model that describes the propagation of a bi-spinor consisting of an electron and a nuclear spin through the network. In addition to a usual random phase contribution at each link, there is a spin-orbit scattering on the links that acts only on the nuclear component of the bi-spinor. In Refs. \cite{BodoHuk,Furusaki} it was shown that  spin-orbit scattering can lead to the appearance of metallic phases in phase diagrams of network models. Our numerical simulations result in similar conclusions. As the Fermi energy crosses the region of delocalized states, there is a smooth change of the Hall conductivity by one unit between the two quantized values. 

Our numerical calculations proceed as follows. 
We study a system of size $M\times L$ where $M$ is the width of the system and $L$ is its length. For a given $M$, the eigenvalues of $\ln (T^+ T)$, where $T$ is the full transfer matrix, behave as $\exp(-2\lambda_nL)$ defining the Lyapunov exponents $\lambda_n$; the smallest positive one, $\lambda_{M/2}$, defines the localization length $\xi_M/M$. The $M$ dependence of
$\xi_M/M$ identifies the phases: (i) a decreasing ratio corresponds to localized state, i.e., an insulator, (ii) a constant ratio corresponds to a critical state, and (iii) an increasing ratio corresponds to a metallic phase. 

The results of our numerical calculations lead to phase diagrams presented in Fig. 
\ref{fig-Phases}. The phase diagram in Fig. \ref{fig-Phases}a corresponds to the spin-degenerate case with zero Zeeman splitting. In the absence of spin-flip scattering ($\delta=0$) there is a transition between the two insulating phases with Hall conductances $\sigma_{xy}=0$ and $\sigma_{xy}=2$ at energy $\epsilon=0$. 
At finite $\delta$, a metallic phase appears in the finite range of energies, as described above. 

The phase diagram for finite Zeeman splitting $\Delta=1$ is shown in Fig. 
\ref{fig-Phases}b.  Without the spin-flip scattering there are two critical energies $\epsilon_c=\pm\Delta/2$ separating the insulating phases with $\sigma_{xy}=0, 1, 2$. A finite spin-flip scattering induces the appearance of two metallic regions around each critical energy, in analogy with Fig. \ref{fig-Phases}a. As the spin-flip scattering rate grows further, the size of each metallic region diminishes and finally collapses to single critical states at energies $\epsilon\approx\pm \Delta/2+\delta/2$. 

The appearance and the subsequent collapse of delocalized phase with the increase of the spin-flip scattering $\delta$ can be qualitatively explained by the detailed consideration  of the mutual influence of the spin-flip scattering at nodes and mixing of the states on links. Elements of the scattering matrix Eqs. (\ref{flip}) -- (\ref{no-flip2}) describe the scattering of the states with antiparallel spins of electron and nucleus.  
Propagating along the links, those states mediate the indirect mixing of the states $\vert \uparrow_e\uparrow_I\rangle$ and $\vert \downarrow_e\downarrow_I\rangle$ by the mechanism described above. 
A typical sequence of transitions leading to a mixing between $\vert \uparrow_e\uparrow_I\rangle$ and $\vert \downarrow_e\downarrow_I\rangle$ looks like $\vert \uparrow_e\uparrow_I\rangle\rightarrow \vert \uparrow_e\downarrow_I\rangle$ on the link, 
$\vert \uparrow_e\downarrow_I\rangle\rightarrow \vert \downarrow_e\uparrow_I\rangle$ at the node, $\vert \downarrow_e\uparrow_I\rangle\rightarrow \vert \downarrow_e\downarrow_I\rangle$ on the link. 
 This mixing acts similarly to the spin-orbit interaction for the {\em electron} wave function and, if effective, results in the appearance of finite regions of delocalized stated around the critical energies $\epsilon_{\uparrow}$ and $\epsilon_{\downarrow}$. The effect of the mixing is reduced with the separation between the energies 
$\epsilon_{\uparrow}$ and $\epsilon_{\downarrow}$, and $\epsilon_1$, $\epsilon_2$. 
For small values of $\delta$ there is a finite interval of energies $\epsilon$, in which the transmission amplitudes $t(\epsilon_1)$ and $t(\epsilon_2)$ are of the same order. In this interval, the effective spin-orbit interaction results in the metallic regions around the critical energies corresponding to the states $\vert \uparrow_e\uparrow_I\rangle$ and $\vert \downarrow_e\downarrow_I\rangle$. At large $\delta$, the large energy separation between $\epsilon_1$ and $\epsilon_2$ leads to the localization of at least one of the states corresponding to those two energies for any $\epsilon$. As a result, the mixing on the links necessarily involves a localized state and hence becomes ineffective. The states $\vert \uparrow_e\uparrow_I\rangle$ and $\vert \downarrow_e\downarrow_I\rangle$ become two decoupled critical states, the metallic regions collapse 
(see Fig. \ref{fig-Phases}). 

In conclusion, we derived the exact solution of spin-flip scattering problem in quantum Hall regime, and used it to construct the network model that describes integer QH transition in presence of scattering by nuclear spins. With increase of the exchange coupling between electron and nuclear spins, we predict the appearance of metallic regions in the QH phase diagram that subsequently collapse to the critical lines. Those critical lines correspond to a standard QH transition.  Our investigation can also shed light on the QH transition in presence of scattering by magnetic impurities.

\begin{acknowledgments} 
The work was mainly done during the authors' stay at MPIPKS, Dresden.  
Authors acknowledge financial support from DFG through
Sonderforschungsbereich 668 and of the SCE internal research grant. 
\end{acknowledgments}

\end{document}